# Time-Domain And MultiMessenger Astrophysics Communications Science Analysis Group Report

Jamie A. Kennea (*Co-Chair*, The Pennsylvania State University)
Judith L. Racusin (*Co-Chair*, NASA Goddard Space Flight Center)
Eric Burns (Louisiana State University)
Brian W. Grefenstettte (California Institute of Technology)
Rebekah A. Hounsell (University of Maryland Baltimore County, NASA Goddard Space Flight Center)
C. Michelle Hui (NASA Marshall Space Flight Center)
Daniel Kocevski (Marshall Space Flight Center)
T. Joseph W. Lazio (Jet Propulsion Laboratory, California Institute of Technology)
Stephen Lesage (Department of Space Science and Center for Space Plasma and Aeronomic Research, University of Alabama in Huntsville)
Tyler A. Pritchard (University of Maryland College Park, NASA Goddard Space Flight Center)
Aaron Tohuvavohu (University of Toronto)
John A. Tomsick (Space Sciences Laboratory, University of California, Berkeley)
David Traore (ORBIT)
Colleen A. Wilson-Hodge (NASA Marshall Space Flight Center)

July 2024

# Executive Summary

The Time-Domain And MultiMessenger (TDAMM) Communications Science Analysis Group (TDAMMCommSAG) was formulated to describe the unique technical challenges of communicating rapidly to and from NASA astrophysics missions studying the most variable, transient, and extreme objects in the Universe. This report describes the study of if and how the transition from current NASA-operated space and ground relays to commercial services will adequately serve these missions. Depending on the individual mission requirements and Concept of Operations (ConOps), TDAMM missions may utilize a rapid low-rate demand access service, a low-rate continuous contact service, low-latency downlink upon demand, or a higher-latency but regular relay service. The specific implementations can vary via space relay or direct to Earth, but requires flexibility and adaptability using modern software infrastructure.

The study team reviewed the current state of NASA communications services and future commercial and NASA communications services under study and in development. We explored the communications capabilities driving from the behavior of the astrophysical objects themselves.

The TDAMMCommSAG report explores the scientific and operational advantages and communications challenges of non-LEO orbits. Non-LEO orbits have significant advantages to TDAMM missions, due to the increased available field of regard, lower radiation environment, and the longer integration times available due to reduction in Earth occultations. However, these orbits also come with significant communications challenges, due to distances involved and the lack of any type of TDRS-like service that can service them. The report finds that this is the main area where NASA could focus on development where the commercial sector has fewer applications.

We explored how TDAMM missions have unique needs for bandwidth, latency, and coverage. TDAMM science drives needs to low latency communications, both reporting events in real-time and commanding to perform rapid observations. High bandwidth downlinks are needed as datasets naturally grow, often beyond the ability to perform onboard processing, and the coverage provided by TDRS now should be extended to future solutions.

We suggested ways that availability and scheduling of communications services would enable TDAMM science. To enable rapid response to transients and variable phenomena, TDAMM missions quickly revise their communications schedules utilizing communications resources with sufficient availability to do so. The interfaces to carry out this scheduling is most efficient if as automated as possible, even without humans in the loop.

TDAMM missions are disadvantaged during the proposal and development stage due to higher communications costs, and we suggest ways this could be mitigated. This includes increasing the availability of costing details to allow proposal teams to better estimate costs during proposal development. We also find that proposal teams should be allowed to do business with commercial providers in order to minimize cost in a competitive environment. However, overall we find that due to high communications needs, TDAMM missions are disadvantaged compared to other mission types due to the associated high costs, and removing communications costs from mission budgets would ensure that mission designs are scientifically motivated rather than communications cost constrained.

As NASA transitions from the current services to wider use of commercial communication services, careful transition planning is essential to provide continuity for current missions, missions in development, and those proposing in the intervening years.

The TDAMMCommSAG aims that this report addresses why TDAMM science uniquely drives communications infrastructure, and why meeting the needs of TDAMM missions will have wider benefits to all NASA missions. The findings within this report are aimed at enabling TDAMM science, from proposal development to operations phase.

# 1 Introduction

Central to all NASA missions are the ability to command our spacecrafts and telemeter data to the ground in an efficient and complete manner. Decades of heavy investment in communications infrastructure are a major reason that NASA missions lead the world in many areas of space science and engineering. The unique access to NASA communications resources enables our missions and their scientific discoveries, but is becoming increasingly more challenging as that infrastructure ages and our instruments become more data intensive.

Like the commercial launch, commercial crew, and commercial cargo programs before it, NASA is moving to rely more heavily on commercial communications systems. Companies providing both space-to-space (relay) and space-to-ground (direct) communications are becoming more plentiful and capable, and hopefully with that, more affordable, available, and flexible.

NASA's Tracking Data Relay Satellite (TDRS) system has been a unique resource for more than 40 years, providing low-latency and high-bandwidth services for missions in LEO. NASA's decisions [1,2,3] to not replenish the TDRS constellation, to gradually deprecate TDRS services, and to transition to commercial space relay providers have major impacts for currently operating missions and those in near-term development. These decisions have caused an uncertain landscape in which to plan and propose future missions, and may lead to the loss of critical capabilities for currently operating missions.

NASA's space-ground Direct-to-Earth (DTE) services are often oversubscribed and are not easily adaptable to meet the needs of missions in terms of rapid scheduling, growing bandwidth needs, and cost.

The Astro 2020 Decadal Survey proclaimed the highest-priority sustaining activity for space to be to "maintain and expand space-based time-domain and follow up facilities in space" [1]. The since-termed Time-Domain And MultiMessenger (TDAMM) astrophysics fleet are missions and observatories that provide multiwavelength and multimessenger observations, often requiring rapid autonomous reactions to the transient Universe. Many astrophysical sources vary with time, but TDAMM enables the study of the evolution of compact objects (black holes, neutron stars, white dwarfs) on all scales.

TDAMM missions have unique communications challenges involving the need to rapidly disseminate transient alerts from space to the ground, rapidly communicate to command either unexpected data downlinks or repointing of observatories, and the need for large datasets to be downlinked quickly to search for transients or characterize rapidly evolving sources. **Table 1** describes the types of communications capabilities that address these communication functions and the associated challenges.

Examples of TDAMM missions that exemplify the use of NASA communications services are the Neil Gehrels Swift Observatory (Swift), with its autonomous detection and re-pointing in response to new transients, and the Fermi Gamma-ray Space Telescope (Fermi), with its all-sky surveys and rapid alerting of transient detection. Both missions are heavy users of TDRS and are great demonstrations of significant scientific discovery enabled by NASA communications systems. TDAMM missions are drivers for improvements to space communications coverage, bandwidth, and scheduling agility, but their needs are not unique. By addressing the needs of TDAMM missions, NASA will address the needs of most other missions.

The TDAMMCommSAG study was conducted in an open and inclusive manner, with welcome participation from members of the community, specific solicitation of scientific leadership from current and upcoming astrophysics missions, and co-chaired by scientists who both study transients, have leadership roles in active and proposed NASA TDAMM missions, and operate key TDAMM infrastructure. The study began in June 2023, lasting approximately 1 year, with regular monthly discussion meetings. This report was written at the conclusions of those discussions by the study co-chairs with input from many members of the SAG and broader community. All co-authors on this study have opted in after reviewing the completed report.

The TDAMMCommSAG evaluated many areas of scientific drivers, communications infrastructure, and programmatic implementation. This report provides detailed findings on ways in which NASA could improve and build a better communications landscape in the coming years for both current and future TDAMM missions.

## 2  Communications Infrastructure

TDAMM missions, and all of the missions for NASA's Science Mission Directorate (SMD), are enabled currently by two complementary communication networks overseen by the Space Communications and Navigation (SCaN) Program: the Near Space Network (NSN) and the Deep Space Network (DSN) (**Figure 1**). The NSN provides services using DTE and Space Relay (SR) assets. NSN's DTE services are provided by both government-owned and commercial ground stations. The NSN's Space Relay services are presently provided using the government-owned TDRS system, with plans to transition to commercial relay providers in the coming years. NASA missions are usually required to either SCaN assets or to obtain commercial services through SCaN, although in some cases purchasing commercial services directly is possible. New commercial and NASA-developed communications resources are coming that must meet the needs of TDAMM future missions.

---

[1] https://spacenews.com/39828nasa-wants-laser-communications-for-tdrs-follow-on-needs-industry-money/
[2] https://spacenews.com/tdrs-launch-marks-end-of-an-era/
[3] https://www.nasa.gov/news-release/nasa-industry-to-collaborate-on-space-communications-by-2025/

| Communication Capability | Usage | Technical Challenges | Examples |
| --- | --- | --- | --- |
| Continuous Contact | Forward - commanding | Spacecraft Power | Space-to-Space Laser Comms |
|  | Return - alerts | Availability, Geographical Coverage, Network Congestion, Complexity (pointing, handovers, etc.) | Scheduling frequent passes on TDRS-MA, DTE, or commercial systems |
| Demand Access | Return - small alerts (e.g. GRB notice). Spacecraft initiated contact. | Availability, Geographical Coverage, Complexity, Additional Hardware | TDRS-DAS, Commercial space communications |
| Low-Latency Data Downlinking | Frequent pre-scheduled passes. On-the-Fly scheduling of passes with low latency | Availability, Geographical Coverage, Automation. Flexibility. | KSAT-lite and other GSaaS providers. |
| High-Latency Data Downlinking | Download bulk data | Latency | DSN, DTE, TDRS-SA |

**Table 1:** A broad range of communications capabilities are needed to meet the requirements of TDAMM missions at both predictable and unpredictable times. While existing NASA resources can meet these needs, they do not necessarily provide adequate flexibility, availability, and cost effectiveness. Upcoming replacement commercial services have the opportunity to provide these services and improve upon current capabilities to maximize the science potential of NASA's future fleet.

## 2.1 Current Communications Systems

NSN is the spacecraft communications and navigation infrastructure for missions within 1 million miles from Earth, covering the low Earth orbit (LEO), geosynchronous orbit (GEO), high Earth orbit (HEO), lunar and Sun-Earth Lagrange point orbital regimes. NSN operate the DTE government-owned ground stations, the TDRS space relay system, and interface to commercial DTE providers. NSN serves missions throughout their entire lifecycle, providing requirements analysis, spectrum management, communications analysis, service agreements, mission design, mission planning, launch, operations, and post-mission support activities. It is used by many current Astrophysics missions including Swift, Fermi, Hubble, NuSTAR, and IXPE.

DSN is the spacecraft communication and navigation infrastructure for NASA's deep space missions. The DSN also presently serves NASA exploration and science missions in lunar and Sun-Earth Lagrange orbits due to current NSN capacity constraints at these distances. DSN consists of three complexes, approximately equally separated in (terrestrial) longitude, with one 70 m diameter radio antenna and multiple 34 m diameter radio antennas at each complex. All antennas are equipped with cryogenic microwave receivers, at a variety of frequencies, for receiving communications from spacecraft and are equipped with high-power transmitters for transmitting commands to spacecraft. Historically, the DSN has been used relatively little by Astrophysics missions, with missions from the Heliophysics and Planetary Sciences Divisions being much more reliant upon it. However, there are notable Astrophysics missions with a TDAMM aspect to their science programs that have used the DSN or are using it currently, including the Chandra X-ray Observatory, the James Webb Space Telescope (JWST), the Transiting Exoplanet Survey Satellite (TESS), and the European Space Agency's Gaia mission.

With their relatively large diameters and sensitive receivers, the DSN antennas are capable of enabling high data rates and high data volumes. However, historically, the DSN has employed a process that schedules which antenna is used for which spacecraft weeks to months in advance. This scheduling process has been driven in part by the need to accommodate Planetary Science missions, which can have limitations due to factors such as a spacecraft being occulted by a planet during a portion of its orbit or a planned close encounter to a planet's satellite. Consequently, the DSN is well suited for large data volume transfers, but the current scheduling process cannot handle rapid minutes-hours scale adjustments, such as are often required for Target of Opportunity observations (ToOs).

Radio astronomy antennas have many similar characteristics as DSN antennas, most notably large diameters and sensitive microwave receivers operating at the required frequencies. There is a long history of radio astronomy antennas being used to augment the DSN capabilities in order to acquire larger data volumes or provide redundancy or both.

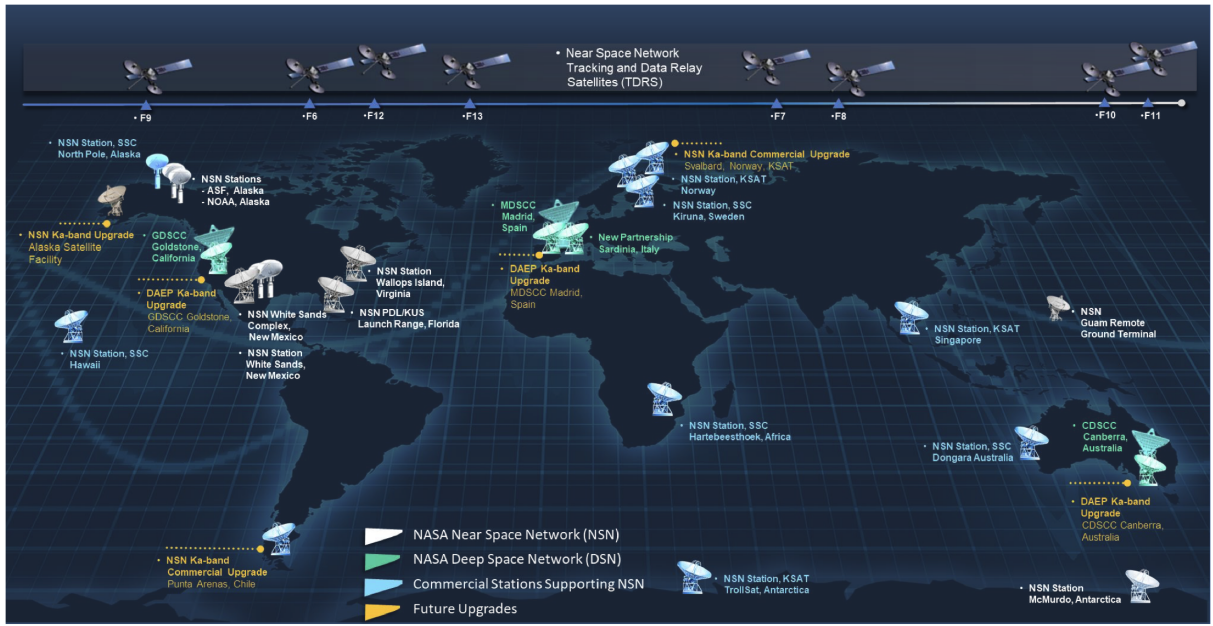

**Figure 1:** Together the DSN and NSN provide ground and space communications for NASA missions from LEO to the Moon and beyond. This map shows the ground and space distribution of communications dishes on Earth and in Earth's orbit. (Credit: https://ntrs.nasa.gov/api/citations/20210020202/downloads/NASACommercialStrategy_WidebandandStds_KaBand2021_R8921.pdf )

Notable examples include the participation of the Parkes Radio Telescope in Voyager 2's flyby of Uranus [2], the participation of the Very Large Array (VLA) in Voyager 2's flyby of Neptune [3], the use of the Murriyang Parkes Radio Telescope in confirming that Voyager 2 had crossed into interstellar space,[4] and a recent experiment in which data were received from the New Horizons spacecraft by the VLA [4]. The scheduling of radio astronomy antennas often is much more accommodating than the DSN for ToOs and other rapid changes in schedule. However, radio astronomy antennas are not equipped with transmitters, so they cannot send commands to spacecraft. For the purposes of receiving either large data volumes or particularly critical data, radio astronomy antennas are feasible augmentations to the DSN's capabilities.

## 2.2 Future Communications Systems

NASA is developing, demonstrating and evaluating several capabilities that offer exciting new possibilities for future TDAMM missions.

### 2.2.1 Communications Services Project

NASA is currently conducting the Communication Services Project (CSP)[5] study of commercial space relay replacements for the TDRS system. TDRS is continuing to support current and shortly upcoming missions, but no new TDRS spacecraft will be launched and services are gradually degrading[6]. CSP was established by the SCaN Program Office to meet NASA's need to migrate near-Earth missions from Government-owned to commercial space-relay communication networks. CSP was chartered to leverage existing commercial infrastructure, capabilities, and market strategies, prompting the formation of public-private partnerships with multiple providers that may eventually be able to offer commercial SATCOM services. CSP is working with these partners to develop and demonstrate capabilities that will extend their existing or planned terrestrial services and support a new space-based user market. In 2022, CSP awarded six Funded Space Act Agreements (FSAA) to commercial industry partners with the objective of demonstrating end-to-end services to meet multiple NASA mission use cases. CSP aims to deliver operationally ready commercial space relay communications services to the NSN by 2030.

---
[4]https://www.nasa.gov/news-release/nasas-voyager-2-probe-enters-interstellar-space/
[5]https://www.nasa.gov/directorates/somd/space-communications-navigation-program/communications-services-project/
[6]https://www.nasa.gov/general/nasas-tracking-and-data-relay-satellite-9-reaches-end-of-mission/

In parallel with the FSAA demonstrations, CSP is coordinating with SCaN and the NSN to develop a portfolio of validated commercial services that will be ready for adoption and acquisition by NASA missions no later than 2030. CSP is actively engaging with the NASA mission community and Mission Directorate representatives to develop a list of service requirements commercial networks must meet to adequately address NASA missions' needs. These service requirements will ultimately be scoped to an industry Request for Proposals (RFP), which will be open for all commercial service providers to bid, not just those participating in the FSAA process.

The TDAMMCommSAG co-chairs have been participating in the CSP study as part of the Commercial Services Users Group (CSUG). Members of the CSP study group have also provided technical input to the TDAMMCommSAG study. These commercial services have the potential to enable scientific discovery and meet the needs of the TDAMM community, but only if they are smoothly transitioned into the services offered by NSN, and made cost effective for NASA missions to utilize.

In addition to the space relay services evaluated as part of CSP, there are growing networks of commercial DTE space-ground communications providers, especially expanding in the higher data-rate bands (e.g., X- and Ka-band). These commercial providers offer efficient and flexible queue scheduling interfaces via application programmable interfaces (APIs), but can be costly, especially those that charge for data egress out of their systems.

### 2.2.2 Lunar Communications Infrastructure

Driven by the requirements of the Artemis lunar exploration program, the NSN is overseeing development and commercial services procurement for several new assets to reduce the near-space utilization burden on the DSN. The new assets include a set of 18-meter-class S-, X-, and Ka-band DTE ground stations suitable for missions out to Earth-Sun Lagrange point distances, and a lunar space relay constellation providing coverage of the lunar south pole[7]. With the goal of delivering a mobile-internet-like experience to mission users, NASA and its commercial and international partners have adopted a standards-based network architecture to ensure interoperability, ostensibly and reduced operational complexity for coordinating services from a diverse set of lunar communications assets [8][9]. Such a model reflects the goals for effective TDAMM space communications.

### 2.2.3 Laser Communications

Other promising commercial and Government communications capabilities include laser/optical communications [e.g., 5]. Laser communications offers the potential for high bandwidth transmissions enabling large data volume transfers. NASA's near-space demonstrations of laser communications include the Terabyte Infra Red Delivery LEO small satellite and DTE system [TBIRD, 6], the Laser Communications Relay Demonstration (LCRD) geosynchronous relay [LCRD, 7], and the Lunar Laser Communications Demonstration [LLCD, 8]. In addition, several companies advertise laser communication flight terminals, ground terminals, or both, either as a package for a mission or as a service with commercial ground infrastructure. Further, various commercial near-space constellations have plans to offer laser communications interfaces to users or to incorporate laser communications crosslinks within their constellations[10]. The Deep Space Optical Communications (DSOC) technical demonstration mission on the Psyche Discovery mission, launched in 2023 October, and has demonstrated the capability of laser beyond the Earth-Moon system [9].

While laser communications offers the potential of much larger data volumes from spacecraft, its applicability to any specific mission likely will require a detailed trade study. Among the potential issues to consider would be the robustness of the ground network, and particularly resiliency to cloud cover at any given ground station; the cost of on-board data storage; spacecraft pointing capabilities or the possibility of a gimbaled system and isolation from the spacecraft; and the required mass and power for adding a laser communications flight terminal.

---

[7] https://ntrs.nasa.gov/api/citations/20230013314/downloads/Commerical%20Lunar%20Ka%20Presentation_FINAL.pdf
[8] https://ntrs.nasa.gov/api/citations/20200001555/downloads/20200001555.pdf
[9] https://www.nasa.gov/directorates/somd/space-communications-navigation-program/lunanet-interoperability-specification/
[10] https://spie.org/news/photonics-focus/mayjune-2024/transforming-tech-of-laserfree-space-optical-communications

## 2.3 Enabling TDAMM Astrophysics through Communications

New DTE and space relay capabilities offer exciting options for future missions that will likely meet the needs of TDAMM missions, at least in LEO orbits, but only if NASA's integration of the provider assets is seamless, efficient and cost effective. A diverse set of providers, where the complexity of services is abstracted from the end user, would have significant benefits for TDAMM missions in all orbital regimes. As the private sector is building new communication services that can benefit both commercial and NASA missions, this provides an opportunity for NASA to innovate new communications capabilities that do not yet have broad commercial applications. Although beyond the scope of this study, future communications solutions should not preclude other critical services that have historically been coupled to space communications infrastructure, such as timing, ranging and navigation services. The rest of this report aims at addressing the specific needs of TDAMM missions, and suggests ways that NASA could implement future communications options and issues for consideration.

> **Case Study: Powerful Solutions Unique to Flagship Missions**
>
> The Nancy Grace Roman Space Telescope, launching in 2027, will generate $\sim 1.4$ TB of survey data per day[a] to downlink from a Sun-Earth L2 orbit. The DSN was considered as a solution, but it did not have sufficient capacity or data rates, but is still being used for commanding. To meet their requirements, Roman has priority access to three Ka-band ground stations that provide near-continuous coverage. ESA and JAXA are contributing one ground station each (ESA ground station is built from scratch, the JAXA is upgrading an existing ground station), while the third is an upgrade to a NASA station at White Sands. This model of large dedicated or prioritized ground stations is used by other data-intensive large missions including the Lunar Reconnaissance Orbiter (LRO) and the Solar Dynamics Observatory (SDO). Flagship missions have the budget and resources to support such operations, but this is not scalable to smaller missions.
>
> ---
> [a] https://svs.gsfc.nasa.gov/13667/

## 3 Science Drivers

Scientific discovery is enabled by agile communications solutions. While all space missions require trade-offs in the design of their communications solutions, TDAMM missions have demanding and complex communications needs. This is due to the unpredictability and rapidly time-variable nature of the phenomena they observe and the coordinated aspect of TDAMM science observation strategies. TDAMM missions do not have a monolithic concept of operations, some TDAMM missions serve in purely monitoring roles, rapidly reporting events; while other TDAMM missions perform follow-up observations to localize, characterize, and track the evolution of the source; whereas missions such as Swift perform both monitoring and follow-up roles. These TDAMM missions rely on heavy communications requirements: For monitoring, having demand access systems that provide continuous reporting of events to the ground, for follow-up missions, having both continuous commanding and/or frequent ground station passes to enable low latency follow-up, in addition to frequent data downlinks in order to rapidly retrieve and analyze data.

It is often the case that missions cannot integrate effectively into the TDAMM infrastructure purely due to the design of their communications ConOps. For example a mission that has a ground station pass only once per day is incapable of performing rapid response follow-up, but the mission itself could be capable of much more if more passes were available. Therefore by improving communications infrastructure for TDAMM, it would not only enable more TDAMM missions, it would enable missions whose primary goals are not driven by TDAMM science, to become TDAMM capable.

By addressing the communications needs of TDAMM missions, we move towards solutions that meet the needs of all missions.

The TDAMMCommSAG explored all of the TDAMM science cases presented in the 2022 TDAMM Workshop Report [10] and reviewed the detailed physics measurements enabled by rapid alerting presented in [11] (i.e. GTN SAG Report, Table 2). We roughly quantified each source class or emission component's needs for the volume of data, how quickly alerts need to be transmitted, how quickly datasets need to be downlinked for ground searches, and how quickly follow-up observations need to begin. The specifics of this quantisation depends on the particular instrument and mission requirements, but this allowed us to establish the broad capabilities needed. We found that some science cases drive requirements, while others do not. Rather than providing the full list of science cases and their requirements, we organize them by the needed observational capability, with examples science cases, and their requirements on communications (**Table 2**).

| Capability | Science Examples | Onboard Alert Latency | Ground Search Latency | Follow-up Latency | Data Volume |
|---|---|---|---|---|---|
| Rapid Discovery and Reporting | GRB prompt emission | seconds | minutes | – | small |
| | SGRs | seconds | – | days | medium |
| | Magnetar Giant Flares | seconds | – | minutes | medium |
| | Type 1 X-ray Bursts | seconds | hours | days | small |
| | Supergiant Fast X-ray Transients | seconds | days | days | medium |
| Rapid Follow Up | GRB afterglows | – | – | minutes | small |
| | Neutron Star Mergers (GW trigger) | – | minutes | seconds | small |
| | Nearby CCSNe ($\nu$ trigger) | – | – | minutes | small |
| | Fast Optical Transients | – | – | hours | small |
| | Kilonova | – | – | hours | medium |
| | Novae | – | hours | days | medium |
| | SN Ia | – | hours | days | small |
| | CCSNe ($\gamma$-ray lines) | – | hours | days | medium |
| Ground Searches of Large Datasets | X-ray binaries | – | days | hours | large |
| | Orphan GRB Afterglows | – | hours | hours | medium |
| Variable Sources in Outburst | Blazars | minutes | hours | hours | large |
| | Non-jetted TDEs | – | hours | hours | large |
| | Jetted TDEs | minutes | hours | minutes | large |

**Table 2:** Science cases drive communications capabilities. However, they are highly dependent on the mission and instrument designs, waveband of observation, and evolution of the astronomical source class. Opening these parameter spaces often results in unforeseen discoveries. Data volumes designations roughly correspond to: small - kB to MB, medium - 10's of MB, large - 100's of MB to GB.

Mergers of a binary neutron star (BNS) system or a neutron star-black hole (NSBH) sytem and all of their associated electromagnetic and multimessenger components provide us with one of the most stringent, complex, and also scientifically relevant example cases. The GW170817/GRB 170817A/AT 2017gfo event [12, 13, 14] from 2017 is still to date our best observational example. In this case, the gravitational wave (GW) network detected a BNS, while near simultaneously and independently, Fermi detected the prompt emission of a gamma-ray burst (GRB), rapidly disseminating that information to the ground via TDRS and distributing a notification to the world through the General Coordinates Network[11] (GCN). Rapid follow-up observations to better pin-point the location of the merger began within hours using dozens of ground- and space-based telescopes including Swift. Thanks to significant advanced planning and development, about 16 minutes after the merger, Swift uploaded a set of pointings to tile the best-known position at that time. Although it initially did not include the position of the optical kilonova that was found by the Swopes telescope $\sim$12 hours after the merger [15], and in the subsequent days, weeks, and years, many dozens of ground and space-based telescopes observed this event, including Hubble and Chandra, making important measurements of the kilonova and GRB afterglow seen from this exceptional scientific event.

The GW170817/GRB 170817A example case demonstrates the scientific potential enabled by rapid alerting, ground searching, and follow-up observations. The rate of these events will increase into the future with upgrades to the GW network. In the next generation GW Cosmic Explorer era, early warning of these events could occur up to 15 minutes prior to the merger itself [16], where it will be vital to have the ability to rapidly repoint space assets to view the light show.

The science cases outlined in this section drive the need for maintaining and improving upon communications capabilities or we will lose out on vital science, preventing current and future missions from living up to their scientific potential.

---
[11]https://gcn.nasa.gov

**Without rapid downlinking of onboard alerts in seconds, we lose:**
- rapid association of multiwavelength and multimessenger transients (e.g. binary neutron star mergers and short GRBs);
- the ability to trigger rapid multiwavelength ground- and space-based follow-up observations;
- operational efficiency for planning follow-up observations resulting in longer delays and requiring more observation time.

**Without rapid downlinking of datasets to perform ground searches in hours, we lose:**
- the ability to detect anything but the brightest of transients in low background environments;
- the ability to keep up with the completeness of large data volumes.

**Without rapid follow-up observations within minutes, we lose:**
- the ability to identify transients when they are their brightest, requiring more observational resources and lower success rate;
- the early evolution of many transients which provides clues into their outflow structure and immediate environment;
- improved localizations to trigger further follow up by sensitive telescopes.

# 4 Non-LEO Orbits

Traditionally TDAMM missions have been in LEO; however, there are significant downsides for missions in LEO. Here we address the reasons why future TDAMM missions may prefer non-LEO orbits, and the communications challenges they will present.

TDAMM missions like Fermi and Swift in LEO face significant issues due to the nearby presence of the Earth. The Earth for a LEO mission covers approximately a third of the sky, and missions like Swift have large Earth limb avoidance angles in order to avoid issues of scattered light, and the combined Earth avoidance angle means that at any one time Swift cannot point at >50% of the sky. The Earth is also a significant source of radiation, through atmospheric scattering of gamma-rays, which limits the localization accuracy of Fermi Gamma-ray Burst Monitor (GBM [17]), and the presence of the South Atlantic Anomaly and Van Allen belts, which both lower observational efficiency and long term can damage some instrumentation. The Earth is also the most significant source of thermal radiation for satellites in a LEO orbit, and non-LEO orbits can provide a much more stable and cooler thermal environment.

In addition, LEO orbits come with stringent requirements for de-orbiting, meaning that future LEO missions may have limited lifespans in order to guarantee re-entry within a specified time span. In addition, growing congestion in LEO makes the possibility of satellite loss due to a conjunction more and more likely, or requires a propulsion system increasing costs, making more sparsely populated orbits more attractive.

> **Case Study: Uncertain Capabilities for TDAMM Missions in HEO and Beyond**
>
> HEO (or cislunar) orbits provide a number of advantages over LEO. From a time-domain science standpoint, moving away from LEO removes the obstruction of the Earth (e.g., the Earth blocks less than a few degrees of the sky compared with nearly half the sky in LEO) so that missions have a larger instantaneous field of regard. This provides opportunities for time domain sky surveys that fuel the discovery of new transient sources. This is the approach that will be used by the Ultraviolet Explorer (UVEX), NASA's next MIDEX mission. This also enables rapid follow-up of transient sources of nearly the entire sky.
>
> However, for imaging in the optical and UV missions with large fields of view and correspondingly large numbers of pixels, the discovery of new transients requires downlinking large amounts of data with relatively low latency. This affects mission design and survey strategy, as the surveys may be interrupted so that data may be transmitted to the ground.
>
> A key challenge in the current environment is knowing what ground station resources are available (and will be available) on timescales that stretch out by a decade. Proposals are often reviewed on the ability of existing stations (regardless of NASA's posture for producing additional ground stations), so that the desired direct-to-earth data rates are continually rising while the available resources to bring the data down to the ground remain fixed. As more missions fly that also require relatively low latency to achieve their time domain science goals (e.g., timescales of hours, not days), available resources from NASA may become even more crowded.

Future TDAMM missions looking to improve on the current generation may look to moving into different orbits in order to enable improved TDAMM science, for example the ability to look at nearly the whole sky is enabled by moving the mission to a Lagrange point, and missions that are interested in intensive monitoring could benefit from cislunar orbits that allow longer stares without Earth occultation and a more stable thermal and radiation environment without atmospheric scattering or radiation belts. Indeed future time domain missions such as UVEX are actively moving to non-LEO orbits, and this trend is expected to continue as the benefits are realized.

For transient science, localization can be improved by using time-of-flight delay localizations such as those provided by the Interplanetary Network (IPN). These rely on comparison of missions spaced at large distances, and therefore to be a productive member of IPN, orbits further than LEO are necessary. However, IPN localizations are typically limited by the large latency when downlinking data from missions at larger distances from Earth [11].

However, moving farther away from the Earth introduces significant issues with communications. Firstly, most TDRS-like realtime solutions only work in LEO. This includes commercial replacements of TDRS that are currently being studied. For orbits greater than 1 million miles from the Earth, DSN is currently required; however, DSN is a limited resource with stringent scheduling requirements, meaning that on-the-fly scheduling or continuous scheduling of DSN assets to cover a TDAMM mission are unrealistic.

Linking smaller commercial ground stations providing continuous connection is a possible solution, provided the real-time data volume is small, i.e. rapid alert containing only pertinent data such as timestamp, location, and a light-curve sample, analogous to those data products currently transmitted through TDRS-Multiple Access (MA) or Demand Access Service (DAS) today.

However, in the future we can expect data sizes to increase even for rapid products. This will mean that future missions looking to take advantage of non-LEO orbits may require continuous high-bandwidth communications solutions that are not currently available. Even in the case where rapid low-latency reporting is not required, TDAMM missions often have to have the ability to respond quickly to ToO requests, which means that either the number of pre-planned uplink passes has to be large to ensure low-enough latency, or we need the ability to rapidly schedule an uplink pass on-the-fly.

Solutions to this problem are not clear. As demonstrated by the NASA's Psyche mission, long distance laser communications may provide a method for obtaining high bandwidth connection from large distances; however, significant investment both in technology and ground infrastructure will be needed in order to generalize this solution. In addition, more ground station infrastructure both commercial and NASA owned may help.

Currently communications to cislunar or Earth-Sun Lagrange point orbits are not an area of significant commercial investment, as the commercial uses of those orbits are undeveloped, and the needs of science missions alone are not likely to spur commercial investment, without NASA funding those efforts. Therefore, communications for non-LEO orbits is a clear area where NASA could invest in developing innovative new communications techniques, leaving the mostly solved problem of communications in LEO to commercial providers.

> **Finding #1: Developing services to provide high-bandwidth, low-latency communications to non-LEO orbits is essential to enable future TDAMM missions.**
>
> **Finding #2: Investment in developing new technologies that can provide spacecraft initiated demand access service or continuous communication links to non-LEO orbits is essential to enable future TDAMM missions.**

## 5 Bandwidth, Latency, Coverage

It is clear that TDAMM science imposes unique constraints communications needs, stressing three key aspects, i.e. that communications provides sufficient bandwidth to obtain large datasets in a timely manner; the need for rapid reporting of events, rapid commanding and rapid downlink of data drives strict latency requirements, whereas the need for concepts such as continuous commanding and alerting drive the need to maximise coverage by communications assets.

**Bandwidth:** As instrument complexity increases, along with increases in the sensitivity of instruments leading to more events being detected, the size of datasets increases with time. Due to limitations in communications infrastructure, the solution to this typically has been some pre-processing of data or filtering of data onboard, to reduce the amount of data being downlinked. This has two obvious downsides: Firstly science may be lost by filtering data, simply because as a mission evolves, new and advanced methods of analysis may be developed that require access to the

> **Case Study: Communications Limit Discovery Potential**
> The Compton Spectrometer and Imager (COSI) is a Small Explorer (SMEX) gamma-ray mission planned for launch into LEO in 2027. COSI is a Compact Compton Telescope (CCT) operating in the 0.2 MeV–5 MeV bandpass with a large (>25%-sky) field of view. Data analysis for COSI (and CCTs in general) is complex and takes substantial computing power. For each gamma-ray, the most likely sequence of Compton scattering interactions needs to be found by carrying out a probabilistic determination of the likelihood of all possible sequences. One of COSI's requirements is to detect and localize GRBs and report their positions in <1 hr (<15 min. goal) of the GRB detection. The COSI team found that the best way to achieve this is to send as much GRB data to the ground as possible after detection and then to process it on the ground using a powerful computer. COSI plans to use TDRS, and given the downlink speed and the fact that the message may need to be repeated multiple times to reach the ground, the COSI team concluded that they will be able to downlink 500 kbits of data per GRB, corresponding to about 1000 gamma-rays. While this is sufficient to meet COSI's basic requirements, the relatively small amount of data limits the positional accuracy that will be achieved, and even if a very bright GRB is detected, it will not result in a more accurate position until the data can be telemetered in the next ground station pass. Concern over reduced TDRS capabilities and coverage has already led to changes to the COSI requirements to account for the possibility that the data may not be successfully downlinked a significant fraction of the time.

raw data. This has indeed happened with two GRB-detecting instruments that have significantly improved the science return, Swift Burst Alert Telescope (BAT [18]) with the *Gamma-ray Urgent Archiver for Novel Opportunities* (GUANO [19]) system, and Fermi-GBM with continuous time-tagged event data [20, 21], enabled by the use of communications which exceeded requirements placed before launch. Secondly, onboard processing of increasingly complex and large datasets requires development of increasingly complex onboard software and increasingly powerful flight computers. Development of flight software has been inherently complex, expensive and risky, and increases in processing power for spacecraft have been slow to arrive. The solution to both problems is to get all data to the ground, and process it there, where processing power is not limited and updates to software do not require such rigorous checks.

**Latency:** The era of TDRS brought us the ability to have realtime alerting from missions through the DAS, as well as rapid commanding. GRB missions such as Swift and Fermi have leveraged this to alert the community essentially in realtime when a GRB is detected. The most extreme example of this was the detection by Fermi-GBM of a sGRB 1.7 s after the GW event GW170817, heralding a new era of TDAMM astronomy. It is clear that the realtime reporting of this event strongly motivated follow-up by many other missions, and allowed unique early observations by other observatories. In addition, and related to the bandwidth discussion above, getting large datasets to the ground for rapid processing is important, meaning that latency requirements are not simply met by having a TDRS-DAS like system. For non-LEO orbits, this situation becomes more complex, as low latency can only be achieved by having regular ground stations contacts.

> **Finding #3:** To support TDAMM Science, future communications solutions should look to ensure that low latency, high bandwidth and high coverage is available for all missions profiles.

**Coverage:** Intrinsically linked to both latency requirements and bandwidth, coverage is simply related to the amount of time that communications are possible, for given parameters. For example, a system like TDRS-DAS aims to provide near 100% availability for GRB alerts by ensuring adequate coverage (as LEO missions transit the three TDRS service regions) and capacity (by pre-allocating receivers on the ground); however, that was interrupted from May 2023 until June 2024 due to the destruction of the Guam ground station by Typhoon Mawar, which left a gap in TDRS coverage for alerts for missions that relied on it. For ground stations, adequate coverage means having sufficient number of ground stations available with the correct capabilities that bandwidth and latency requirements can be met.

The TDRS system that needs to be replaced by commercial providers has been providing all three capabilities needed for TDAMM missions. SA provides high-bandwidth downlink with high coverage, utilized by missions such as Hubble and Fermi. DAS provides low-latency alerts with high coverage (near 100% pre-Guam damage). MA Forward provides low latency commanding. A TDRS replacement could provide all these capabilities (although it is not clear if a replacement for SA passes will be provided); however, this only covers LEO, and providing low-latency, high-bandwidth and high-coverage communications at non-LEO orbits is extremely challenging, given the current over-subscription and availability of ground assets such as DSN that can communicate outside of LEO. So far these plans have typically involved providing TDRS-like coverage using continuous ground station passes, but this can incur a high cost. Therefore solutions to provide low-latency, high-bandwidth and high-coverage communications need to be put in place for both LEO and beyond LEO.

> **Finding #4: The flexibility of TDRS has enabled TDAMM science, therefore access to similar solutions with commercial services in the future will be crucial.**
>
> **Finding #5: Due to the limitations of onboard computing, some missions require rapid or high cadence downlinks of large datasets to detect and characterize transients - a function that is not currently possible.**

# 6 Availability & Scheduling

TDAMM sources go into outburst on unpredictable schedules and the timescales drive the need for flexibility in scheduling communications for both forward commanding and data return. The current model for NSN and DSN scheduling poses challenges for TDAMM observations. NSN and DSN are typically oversubscribed, especially when it comes to periods around launches, spacecraft emergencies, and the priorities of flagship missions where network capacity can become a limitation. Typically TDAMM missions are smaller (CubeSat, SmallSat, Explorer, Probe) and therefore lower on the priority lists. These availability and scheduling issues may be alleviated with future commercial resources, but those capabilities need to be cost effective.

> **Finding #6: High availability of communication networks on short notice is essential for TDAMM missions to rapidly schedule communications assets to both respond to target of opportunity (ToOs) follow-up observations and prioritize downlinking of data around events of interest.**

Scheduling NSN and DSN DTE assets requires labor-intensive iterative schedule generation for each mission months or weeks in advance, with updates to those schedules as passes approach and orbit determination is improved though interaction with human schedulers rather than software interfaces. By contrast, NSN's TDRS space relay uses the Space Network Access System (SNAS) for communications scheduling, which provides a software interface for programmatic scheduling and is adaptable for rapid ($\sim$15 minute) revisions to scheduling.

NASA underwent the Space Network Ground Segment Sustaining (SGSS) project from 2014 - 2023 to modernize the ground infrastructure for the SN including scheduling tools to "implement a flexible and extensible ground segment that will allow the ... SN to maintain the high level of service in the future, accommodate new users and capabilities, while reducing the effort and cost required to operate and maintain the system"[12]. The cancellation of the project in 2023 resulted in sustained use of the existing scheduling tools.

Future commercial capabilities of on-the-fly scheduling through efficient Application Programming Interfaces (APIs; e.g. AWS[13]) or low-bandwidth constant connectivity may offer elegant solutions to the needs to react to a new astronomical transient by the NASA TDAMM fleet. Queue scheduling and automation are needed to efficiently react to the discoveries and rapid-response required for the next generation TDAMM missions.

Currently operating commercial providers offer APIs and scheduling interfaces; however, SCaN does not make these services available to NASA missions, rather making users go through it's own interfaces. Missions should be able to use those commercial interfaces, as providing these services through NSN may require another extensive interface development project like SGSS. Moving to commercial providers should mean that infrastructure for scheduling is improved, made more efficient and easier to use, not made more burdensome for the mission teams.

> **Finding #7: The use of efficient modern commercial scheduling interfaces (e.g APIs) will enable TDAMM observations, and are more efficient than existing SCaN interfaces. APIs provide realtime view of availability and eliminate back and forward interactions with human schedulers.**

---

[12]https://gdmissionsystems.com/satellite-ground-systems/space-network-ground-segment-sustainment
[13]https://docs.aws.amazon.com/ground-station/latest/APIReference/Welcome.html

> **Case Study: Frequent Communications for Large Datasets is Not Unique to TDAMM**
> Even for missions where time-domain science is not primary, communications, commanding, and data transfer from the spacecraft to the science teams remains a major factor in the science return on NASA's investment in space instrumentation. Survey missions such as SPHEREx require frequent communications to keep up with their large data volumes. This is especially true for missions in LEO, where the available ground station networks may only provide infrequent opportunities for high bandwidth data transfer. As demand for bandwidth grows, it is not clear that the existing infrastructure is sufficient to address even the non-TDAMM question of "How does a mission recover all of their data?". As stated above, data demands from *all* missions continue to rise, while the availability of resources to bring the data down remains flat.

# 7 Cost

Due to stringent communications requirements driven by low latencies and high coverage needed for TDAMM missions, costs for TDAMM missions can often be much higher than non-TDAMM missions. At the proposal phase, this places TDAMM missions at an immediate competitive disadvantage compared to other missions that do not have such stringent communications requirements, as high communications costs inevitably take money away from instruments and operations. For example missions that require rapid reporting of transient events or rapid response to external triggers, may require multiple communications solutions in order to meet science requirements, due to the need to support rapid reporting and commanding through a TDRS analogous space network, and ground station support for higher bandwidth requirements of downlinking data. Cost limitations can lead missions to reduce the instrument resolution (temporal, spectral, spatial) to manage communications leading to a loss of science. If the science requirements of a mission requires low latency access to data, this can often only be solved by simply having ground station passes more frequently, increasing costs.

Therefore, we wish to highlight three key areas to help ensure that TDAMM missions are not placed at a competitive disadvantage to non-TDAMM missions during an Announcement of Opportunity (AO) call:

**Cost openness:** Currently, it is typical to require proposers to contact SCaN in order to design and determine cost of communications solutions for a mission. However, for missions in development this is a task that could often require a rethink or redesign as the design and budgeting process evolves during proposal development. This leads to proposers often waiting until the last minute to contact SCaN, which in turn places pressure on SCaN to quickly turn around results in order to meet proposer deadlines. As up-to-date numbers and details are which are not easily accessible are needed to estimate costs, home grown cost estimates can often differ from what SCaN produces.

In the past there has also been a lack of responsiveness on the part of SCaN, or more egregiously, SCaN has sometimes given incorrect information to missions teams. These issues affect both mission development and in some cases, can create major problems at the review stage, at which point it is often too late to fix, leading to the mission ending up with a high risk rating, affecting selectability.

These issues could be avoided by making the tools and details required to build accurate self-assessed cost estimates available to teams, without relying on SCaN, and ensuring that these tools cover all use cases. Instead, SCaN could provide validation of costs before proposal submission, and support to mission teams during proposal development.

We note that DSN already provides such a tool for cost estimation, the DSN Aperture Cost Calculator[14], providing a model for how cost estimation tools could be provided for other NSN assets. However, development of complex web tools may not be necessary if adequate, easy to locate, and up-to-date documentation is made available to teams to self-assess costs.

> **Finding #8: In order to ensure that reliable and timely cost estimation by proposal teams is possible, teams need easy access to up-to-date documentation or tools to allow an accurate self assessment of projected communications costs.**

**Cost competitiveness:** The introduction of commercial providers should allow for more competitive pricing to be realized, as mission teams themselves could request different communications providers to competitively bid on contracts which provide communications support. However, in current AOs (e.g., the 2023 Astrophysics Explorers Probe) it has been stated that "When the use of non-NASA communication services is proposed, NASA reserves the option of contracting for those services directly through its SCaN office."[15] This may adversely affect costs as SCaN overrides direct deals with its own cost structure that may not be as competitively priced as a direct commercial quote. Therefore, all attempts to realize commercial cost competitiveness should be made, giving teams the flexibility of either going to commercial providers directly, or going through SCaN.

---

[14] https://dse.jpl.nasa.gov/ext/

**Finding #9: Pursuing direct relationships with commercial communications providers allows proposal teams to realize potential cost savings.**

**Removing communications cost from AOs:** For highly cost constrained missions such as CubeSats, SmallSats and even SMEX missions, heavy communications requirements can become a significant fraction of the PI Managed Mission Cost (PIMMC), leaving less money for hardware, operations, and science. In some cases, communications requirements effectively disallow missions due to the cost burden. Solutions to this problem often include reducing scientific effectiveness of missions to keep communications costs low, including increased latency of response and notification, and reducing bandwidth requirements by doing more onboard filtering of data. One clear way to solve this problem would be to level the playing field on communications by not requiring missions to carry communications costs in their PIMMC, and only requiring that missions well motivate their communications use based on the science case. This approach has been adopted in previous Planetary Science AOs, e.g., the Discovery 2014[16] in order to maximise science without disadvantaging science cases with heavy communications requirements. Additionally, once missions are selected, communication costs are typically not paid directly by the missions themselves, and in some cases missions have increased their communications volume post-launch, and this did not incur any additional costs to the missions teams.

**Finding #10: Removing communications costs from PI Managed Mission Costs would ensure TDAMM missions are not disadvantaged compared to other missions with less burdensome communications requirements, and ensure communications needs are scientifically motivated, rather than cost-driven.**

TDAMM missions and missions in general have often found ways to reduce cost by going to contributed communications infrastructure. One example of this is the Malindi ground station provided by the Italian Space Agency (ASI), and used by missions such as *Swift*, NuSTAR, and IXPE. Although contributed ground stations like Malindi allows for missions to control costs, this is often at the disadvantage of both a more complex communications infrastructure, and congestion at Malindi itself as multiple missions take advantage of this service. It is clear that by reducing or removing communications costs from AOs, this would free missions to choose a solution which best meets the science and technical needs.

---

[15]https://nspires.nasaprs.com/external/viewrepositorydocument/cmdocumentid=953461/solicitationId=
%7B6069A30C-99DE-3647-6AF9-9D5646C9210A%7D/viewSolicitationDocument=1/AstroProbes%20AO%20Amendment1%
202023-09-08.pdf

[16]https://nspires.nasaprs.com/external/viewrepositorydocument/cmdocumentid=438340/solicitationId=
%7BFE7B4C63-873D-63C1-4D15-1D46E2FEA949%7D/viewSolicitationDocument=1/discovery-2014-amend1.pdf

---

**Case Study: Continuous Communications Unattainable for Small Missions** For TDAMM missions, real-time alerts and/or fast data downlink is crucial for the science and for enabling follow-up observation. This drives the need for continuous coverage and requires a minimum of three ground stations distributed across the globe. If the data volume is low (low latency alert with timestamp and location), it could be achieved by small dedicated commercial antennas which normally require a contract of minimum operation period in excess of $1M per year (FY21$). From the document Space Communications and Navigation Mission Operations and Communications Services (SCaN-MOCS-0001, Revision 4), the rates for cost estimate of the SCaN networks is $550 per pass and each pass is $\leq 30$ minutes for NSN Direct-to-Earth assets (Table 5-2). To achieve continuous coverage for TDAMM mission, this translates to an operational cost of $9.6M per year on communication alone. Commercial providers such as KSAT are now available via SCaN, and for reduced data volume that can utilize smaller aperture ground stations, the cost can be lower at $110 per 30-min pass (SCaN quote from 2021 Astrophysics Explorer call). Even with this lower quote, annual TDAMM mission operation cost on real-time alerts alone is $1.9M per year. Any additional cost saving measured by increasing latency would negatively impact science return due to the nature of TDAMM science.

# 8 Transition Planning & AOs

The decommissioning of TDRS, with no clear path forward means that TDAMM missions in development (e.g., COSI) or at the proposal stage (e.g., APEX Probes), are required to "roll their own" TDRS-like communications solution. This uncertainty in the future of communications has greatly disadvantaged TDAMM missions, with no currently NASA-selected path forward. This is driven by the fact that access to TDRS by future missions is cut off long before replacements were available. The communications requirements in the AO could have ameliorated this issue by simply stating that missions requiring TDRS-like communications should simply baseline TDRS, but plan to use one of the selected commercial vendors, although this would firmly place $1B NASA missions in a position of being "early adopters" of these upcoming commercial services.

This highlights the need for better transition planning when making a major service change like this. Replacements should be in place or at least, expected to be in place by launch for any mission AO currently being proposed. The lack of such services creates a path for missions concepts that rely on TDRS-like service to be considered inherently high risk due to uncertainty created by the transition.

> **Finding #11:** A smooth transition between communications services in the future ensures that AOs do not disadvantage missions whose science case depends on a service which is in transition.

The implementation of communications costing varies across AOs from including them in the PI-managed cost throughout the mission, to carrying communications costs through proposal and then managed outside PI-managed costs for the duration of mission even when the communications needs change, to completely excluding them from the PI-managed costs. These variations have led to miscommunications with SCaN during the proposal stage, ultimately disadvantaging missions with significant communications usage (e.g., TDAMM missions).

> **Finding #12:** Consistent communications requirements and costing across all NASA SMD AOs would ensure more accurate proposal development and evaluation.

The scale of licensing, testing, and operational costs of using future commercial communications options through SCaN or NASA-wide contracts must scale reasonably even if multiple solutions are needed. Commercial space communications solutions must be affordable and operational on the right timescales for missions in development today. If commercial services are to be provided via SCaN, there must be incentives (e.g., bulk discounts, common scheduling interfaces, access to multiple providers). There also needs to be some guarantee that those commercial providers will be able to continue to provide those services at affordable rates for the duration of the extended mission, accepting the reality that extended missions can often continue 10-20 years or more.

> **Finding #13:** Access to multiple providers through a common interface has potential to provide affordability and long-term stability to missions via bulk investment by SCaN.

> **Case Study: Lack of a Transition Plan is Harming Missions in Development**
>
> When COSI was proposed in 2019 and even in Phase A (2020-2021), information about TDRS decommissioning was not available. Some information became available during the first part of Phase B, but there are details that are still being provided. For example, the DAS, which is used by current satellites, may not be available for COSI, requiring more resources from COSI's Mission Operations Center. Although COSI does expect to use TDRS during its 2-year prime mission (ending in late-2029), the degradation of TDRS may accelerate as decommissioning is planned to be underway. With its S-band communications, it is unlikely that COSI will be able to use any commercial space relay options in an extended mission, and COSI may be using only ground station passes for telemetering data.
>
> For the recent 2023 Astrophysics Explorer Probe AO, no clear statement on the plan for supporting TDRS-like communications was given, therefore missions proposing to the Probe AO needed to figure out how to support this. The AO advice was "you cannot baseline TDRS, contact SCaN if you need to use realtime communications, to find a solution." However, STROBE-X reported getting the following response from SCaN when requesting support: "NSN is currently unable to plan or cost out "integration/operational" services for the commercial Earth-Proximity Space-Relay service providers that Glenn is currently evaluating."

# Acknowledgements

We acknowledge discussion from all those who attended the TDAMMCommSAG monthly meetings. Part of this research was carried out at the Jet Propulsion Laboratory, California Institute of Technology, under a contract with the National Aeronautics and Space Administration. Some of this material is pre-decisional information and for planning and discussion only. We gratefully acknowledge contributions to this report from helpful discussions with Christopher J. Roberts of NASA/GSFC, and Aaron Smith and Ryan Richards of CSP / NASA/GRC.

# References

[1] National Academies of Sciences, Engineering, and Medicine. *Pathways to Discovery in Astronomy and Astrophysics for the 2020s.* 2021.

[2] C. D. Bartok. Catching the whispers from Uranus. *Aerospace America*, 24:44–46, May 1986.

[3] J. W. Layland and D. W. Brown. Planning for VLA/DSN Arrayed Support to the Voyager at Neptune. *Telecommunications and Data Acquisition Progress Report*, 82:125–135, April 1985.

[4] Parker White, Robert Jensen, Justin Bradfield, Connor Thompson, and David Copeland. New Horizons VLA Experiment. In *2024 IEEE Aerospace Conference (AERO)*, 2024.

[5] Bernard L. Edwards, Dimitrios Antsos, Abhijit Biswas, Lena Braatz, and Bryan Robinson. An envisioned future for space optical communications. In *2023 IEEE International Conference on Space Optical Systems and Applications (ICSOS)*, pages 1–7, 2023.

[6] Curt M. Schieler et al. On-orbit demonstration of 200-Gbps laser communication downlink from the TBIRD CubeSat. In Hamid Hemmati and Bryan S. Robinson, editors, *Free-Space Laser Communications XXXV*, volume 12413, page 1241302. International Society for Optics and Photonics, SPIE, 2023.

[7] David J. Israel et al. NASA's laser communications relay demonstration (LCRD) experiment program: characterization and initial operations. In Hamid Hemmati and Bryan S. Robinson, editors, *Free-Space Laser Communications XXXVI*, volume 12877, page 1287705. International Society for Optics and Photonics, SPIE, 2024.

[8] Don M. Boroson et al. Overview and results of the Lunar Laser Communication Demonstration. In Hamid Hemmati and Don M. Boroson, editors, *Free-Space Laser Communication and Atmospheric Propagation XXVI*, volume 8971, page 89710S. International Society for Optics and Photonics, SPIE, 2014.

[9] Abhijit Biswas et al. Deep space optical communications technology demonstration. In Hamid Hemmati and Bryan S. Robinson, editors, *Free-Space Laser Communications XXXVI*, volume 12877, page 1287706. International Society for Optics and Photonics, SPIE, 2024.

[10] The TDAMM Workshop Science Organizing Committee. The Dynamic Universe: realizing the science potential of Time Domain and Multi-Messenger Astrophysics (TDAMM).

[11] Eric Burns et al. Gamma-ray Transient Network Science Analysis Group Report. *arXiv e-prints*, page arXiv:2308.04485, August 2023.

[12] B. P. Abbott et al. GW170817: Observation of Gravitational Waves from a Binary Neutron Star Inspiral. , 119(16):161101, October 2017.

[13] B. P. Abbott et al. Gravitational Waves and Gamma-Rays from a Binary Neutron Star Merger: GW170817 and GRB 170817A. , 848(2):L13, October 2017.

[14] B. P. Abbott et al. Multi-messenger Observations of a Binary Neutron Star Merger. , 848(2):L12, October 2017.

[15] D. A. Coulter et al. Swope Supernova Survey 2017a (SSS17a), the optical counterpart to a gravitational wave source. *Science*, 358(6370):1556–1558, December 2017.

[16] B. Banerjee et al. Pre-merger alert to detect prompt emission in very-high-energy gamma-rays from binary neutron star mergers: Einstein Telescope and Cherenkov Telescope Array synergy. , 678:A126, October 2023.

[17] Charles Meegan et al. The Fermi Gamma-ray Burst Monitor. , 702(1):791–804, September 2009.

[18] Scott D. Barthelmy et al. The Burst Alert Telescope (BAT) on the SWIFT Midex Mission. , 120(3-4):143–164, October 2005.

[19] Aaron Tohuvavohu et al. Gamma-Ray Urgent Archiver for Novel Opportunities (GUANO): Swift/BAT Event Data Dumps on Demand to Enable Sensitive Subthreshold GRB Searches. , 900(1):35, September 2020.

[20] C. L. Fletcher, J. Wood, A. Goldstein, and E. Burns. Gamma-ray Follow-up of the LIGO/Virgo Third Observational Run (O3) with Fermi-GBM. In *American Astronomical Society Meeting Abstracts*, volume 237 of *American Astronomical Society Meeting Abstracts*, page 125.07, January 2021.

[21] O. J. Roberts et al. The First Fermi-GBM Terrestrial Gamma Ray Flash Catalog. *Journal of Geophysical Research (Space Physics)*, 123(5):4381–4401, May 2018.

# Acronyms and Abbreviations

**APEX**  Astrophysics Probe Explorer

**API**  Application Programming Interfaces

**AO**  Announcement of Opportunity

**ASI**  Italian Space Agency

**AWS**  Amazon Web Services

**BAT**  Burst Alert Telescope (Swift)

**BNS**  Binary Neutron Star

**CCSNe**  Core Collapse Supernovae

**CCT**  Compact Compton Telescope

**ConOps**  Concept of Operations

**COSI**  Compton Spectrometer and Imager

**CSP**  Communication Services Project

**CSUG**  Commercial Services Users Group

**DAS**  Demand Access Service (TDRS)

**DSN**  Deep Space Network

**DSOC**  Deep Space Optical Communications

**DTE**  Direct to Earth

**ESA**  European Space Agency

**Fermi**  Fermi Gamma-ray Space Telescope

**FSAA**  Funded Space Act Agreements

**GBM**  Gamma-ray Burst Monitor (Fermi)

**GCN**  General Coordinates Network

**GEO**  Geosynchronous Orbit

**GRB**  Gamma-ray Burst

**GTN**  Gamma-ray Transient Network

**GW**  Gravitational Wave

**HEO**  High Earth Orbit

**Hubble**  Hubble Space Telescope

**IPN**  Interplanetary Network

**IXPE**  Imaging X-ray Polarimetry Explorer

**JAXA**  Japan Aerospace Exploration Agency

**JWST**  James Webb Space Telescope

**LCRD**  Laser Communications Relay Demonstration

**LEO**  Low Earth Orbit

**LLCD**  Lunar Laser Communications Demonstration

**LRO**  Lunar Reconnaissance Orbiter

**MA**  Multiple Access (TDRS)

**MIDEX**  Medium Explorer

**NSBH**  Neutron Star-Black Hole

**NSN**  Near Space Network

**NuSTAR**  Nuclear Spectroscopic Telescope Array

**PIMMC**  Pi Managed Mission Cost

**RFP**  Request for Proposals

**Roman**  Nancy Grace Roman Space Telescope

**SA**     Single Access (TDRS)

**SAG**  Science Activity Group

**SCaN**  Space Communications and Navigation

**SDO**  Solar Dynamics Observatory

**SGR**  Soft Gamma-ray Repeater

**SGSS**  Space Network Ground Segment Sustaining

**SMD**  Science Mission Directorate

**SMEX**  Small Explorer

**SN**     Space Network

**SN Ia**  Supernovae Type Ia

**SNAS**  Space Network Access System SR Space Relay

**Swift**  Neil Gehrels Swift Observatory

**TBIRD**  Terabye InfraRed Delivery

**TDAMM**  Time-Domain and Multimessenger [astrophysics]

**TDAMMCommSAG**  Time-Domain and Multimessenger astrophysics Communications Science Activity Group

**TDE**  Tidal Disruption Event

**TDRS**  Tracking and Data Relay Satellites

**TESS**  Transiting Exoplanet Survey Satellite

**ToO**  Target of Opportunity

**UVEX**  Ultraviolet Explorer

**VLA**  Very Large Array